\documentstyle[psfig,12pt,twocolumn]{article}

\begin{document}

\setlength{\textwidth}{480pt}
\setlength{\textheight}{630pt}
\setlength{\topmargin}{-0.375in}
\setlength{\oddsidemargin}{-0.0833in}
\setlength{\evensidemargin}{-0.0833in}
\setlength{\parindent}{0in}
\setlength{\parskip}{6pt}

\title{
Are There ``Neutrino Flavors"? : More on the Equation $H=mv^2$}

\author{
{\large Ezzat G. Bakhoum}\\
\\
{\normalsize New Jersey Institute of Technology}\\
{\normalsize P.O. Box 305, Marlton, NJ. 08053 USA}\\
{\normalsize Email: bakhoum@modernphysics.org}\\
\\
{\normalsize (This work has been copyrighted with the Library of Congress)}\\
{\normalsize Copyright \copyright 2003 by Ezzat G. Bakhoum}
}

\date{\normalsize Posted: Feb. 26, 2003 \hspace*{.25in} Updated: Mar. 25, 2003}

\maketitle

{\large\bf Abstract:}\\
\\
In a recent work by the author\cite{Bakhoum1} it was demonstrated that the ``neutrino" that is currently being detected in solar reactions (and possibly in other high energy nuclear reactions) is not Pauli's particle; as it became clear that such a particle is an unnecessary proposition in explaining the beta-decay reaction. In addition, it became obvious that concepts such as ``flavor oscillations" of the solar neutrinos are superfluous. In this paper it is further demonstrated that if certain key experiments of the past are reinterpreted in view of the total energy equation $H=mv^2$, then the principle of the existence of different neutrino ``flavors" is itself very questionable.
 
\pagebreak

{\large\bf 1. Introduction:}\\
\\
In an entertaining book by the title ``The God Particle" \cite{Lederman1}, distinguished experimental physicist Leon Lederman describes the first experiment in which the different ``neutrino flavors" were detected. He wrote: ``In almost all of our events, however, the product of the neutrino collision was a muon. {\em Our} neutrinos refused to produce electrons. Why? ...somehow ``muon" was imprinted on them". In that 1962 experiment\cite{Lederman2}, Lederman and his co-experimenters demonstrated, by using a spark chamber, that a number of muon events that originated inside the chamber were very possibly due to the neutrino that was thought to be generated in the decay reaction

\begin{equation}
\pi^{\pm} \rightarrow \mu^{\pm} + \nu.
\label{1}
\end{equation}

That reaction was postulated by Lattes et al.\cite{Lattes} in 1947 in order to satisfy Einstein's equation $H=mc^2$, since it was observed that the pion($\pi$) and the muon($\mu$) are particles with different masses. Since the neutrino generated in that reaction was considered to be the prime candidate for the  events observed by Lederman and his colleagues, a number of other important explanations for the 1962 experiment were unfortunately overlooked\footnote{The author regrets that numerous attempts to communicate with professor Lederman before the publication of this paper were unsuccessful.} (this is particularly true in view of recent knowledge\cite{Appel} that wasn't available in 1962).\\ 
\\
In an earlier paper by the author\cite{Bakhoum1} it was demonstrated that a total energy equation that will satisfy the theory of special relativity as well as the fundamental postulates of the wave mechanics introduced by Compton and de Broglie will be the equation $H=mv^2$, where $m$ is the relativistic mass of a particle and $v$ is its velocity. It was shown that while the traditional quantity $H = m c^2$ does emerge in certain experimental cases (such as in electron-positron annihilation), it fails to emerge in many other cases (such as in modern nuclear fission experiments; beta-decay; etc). The conclusion that emerged from that work was that the expression $H=mc^2$ is to be understood as a special case of the broader expression $H=mv^2$, which does explain all the observed experimental facts. The total energy as given by the quantity $mc^2$ is of course obtained only when the velocity of a particle in a given reaction is close to $c$.\\
\\
In this paper we will discuss Lederman's experiment in view of the total energy equation $H=mv^2$ and in connection with three earlier key experiments\cite{Lattes},\cite{Bjorklund},\cite{Steinberger} in which the decay of neutral and charged pions was discovered. This paper is organized as follows: in Sec.2 we will discuss first the 1950 experiments of Bjorklund et al.\cite{Bjorklund} and of Steinberger et al.\cite{Steinberger} that dealt with the decay of neutral pions. It will be demonstrated that the results of those two experiments very strongly support the equation $H=mv^2$, not $H=mc^2$. In Sec.3, the experiment of Lattes et al.\cite{Lattes} in which the decay of charged pions was discovered will be discussed. It will be shown that the data presented by the authors, if interpreted within the framework of the theory of  Quantum Chromodynamics (QCD), it rules out the possibility of the emission of a  neutrino and further supports the total energy equation $H=mv^2$. Hence, it will be proven once again that the neutrino is an unnecessary proposition in explaining the decay of a pion into a muon. Finally, Lederman's 1962 experiment\cite{Lederman2} will be discussed. Given the absence of a neutrino in $\pi^{\pm}$ decay, and given recent experimental findings\cite{Appel} that were likely unknown in 1962, a very plausible interpretation of the result of that experiment will be given. Accordingly, the principle of the existence of a ``muon neutrino" becomes very questionable. Sec.4 summarizes the conclusions of this paper, along with an explanation of why a reaction such as $\mu^{+} \rightarrow e^{+} + \gamma$ has never been observed.\\
\\
\\
{\large\bf 2. The neutral pion ($\pi^0$) decay experiments and further verification of the equation $H=mv^2$:}\\
\\
In a series of two key experiments performed in the period of 1949-1950 at the radiation laboratory at Berkeley, neutral pions were generated and studied for the first time in the laboratory. Bjorklund et al. performed the first experiment. The second experiment was performed by Steinberger et al. and used an improved experimental technique. Both experiments show results from which we can withdraw some very important conclusions about the principle of mass-energy equivalence.\\
\\
{\bf 2.1. The experiment of Bjorklund et al.\cite{Bjorklund}:}\\
\\
In the first experiment by Bjorklund et al., a carbon target was placed inside a cyclotron chamber and bombarded with protons accelerated to different energies. The authors observed the emission of high-energy photons from the target, the spectrum of which does not fit that of a bremsstrahlung.  The explanation was that a neutral meson must be decaying into two photons, as predicted by Lewis, Oppenheimer and Wouthuysen a few years earlier. The results of that experiment were summarized in one figure that was given by the authors (Fig.4a in the authors' manuscript):

\par
\vspace*{.15in}

\psfig{figure=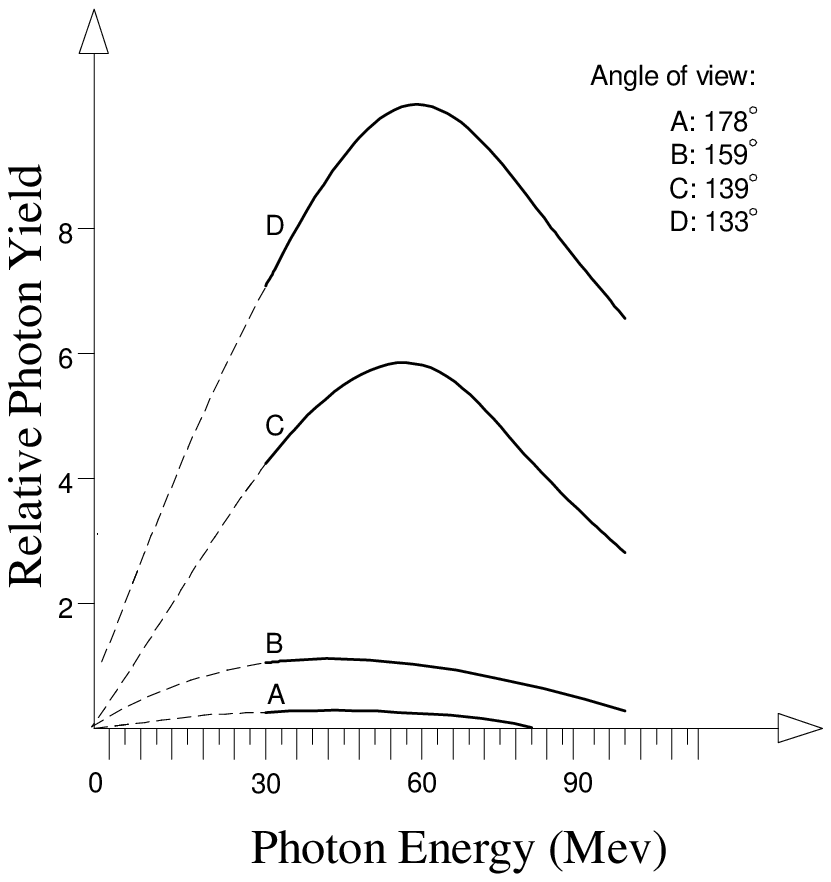}

\vspace*{.1in}

{\small Figure 1: Distribution of the photon energies detected in the decay of neutral pions as given by Bjorklund et al. (the dotted portions of the curves are not shown in the authors' figure). The difficulty of detecting photons at an angle of precisely 180 degrees indicate the impossibility of the decay of a stationary pion. The very wide distribution of the energies of the pions support the total energy equation $H=mv^2$, not $H=mc^2$ (see discussion).}\\
\\
This figure shows the yield of the observed $\gamma$ photons versus the measured energy of the photons. Curve A shows the results when the incident proton energy was 180Mev. Curves B through D correspond to higher proton energies. For each curve, the authors showed the angle at which the photon detector was able to detect the emitted photons (the angles are as indicated on the figure). The curves show some remarkable facts:
\begin{itemize}
\item Observation of the decay photons did not start at an angle of precisely $180^{\circ}$, but at an angle of $178^{\circ}$ instead (curve A). This indicates the impossibility of detecting the decay of a stationary pion (a stationary pion decays into two back-to-back photons emitted at angles of exactly $0^{\circ}$ and $180^{\circ}$). The impossibility of detecting the decay of stationary pions, is, of course, in agreement with the prediction of the equation $H=mv^2$ (since $H=0$ at rest).  We must note here that the energy used to obtain curve A was 180Mev, which was just above the production threshold of the pion (which the authors estimated to be about 175Mev or higher). Therefore it is not inconceivable that some of the pions so produced must be almost stationary; yet the decay of those almost stationary pions wasn't of course observed.
\item The estimated mass of the $\pi^0$ in those early experiments was about 150Mev/$c^2$. The modern estimate (based on a relativistic calculation) is 135Mev/$c^2$. If we take the modern estimate, then we would expect that the energy of each emitted photon to be always higher than $\frac{1}{2}m_0c^2$, or 67.5Mev. Yet, as the curves show, photons of energies as low as 30Mev were observed. The authors argued in their manuscript that this effect is just due to Doppler shift, as a result of the high velocities of the pions. However, it is difficult to make a case for a Doppler shift for two reasons: first, the photon is obviously emitted {\em after} the decay of the particle, not before. Hence, a measure of the total energy of the particle should be expected. Secondly, it is actually a very simple problem of kinematics to prove that if the photon is observed, then the momentum of the decaying particle with respect to the observer must be positive, not negative. Hence, a negative Doppler shift as argued by the authors must be ruled out. The plausible explanation, therefore, is that the total energy equation $H=mv^2$ must be valid. Note that we have showed the ``missing" portions of the curves as dotted lines, since, of course, those portions weren't obtained experimentally (detecting the decay of a slow pion would require a narrower observation angle).  
\end{itemize}

From these experimental facts it is clear that, just as in the case of beta-decay, the spectrum of pion decay is obviously a continuous spectrum. This does not support the mathematical expression $H=mc^2$ (since it has a minimum limit of $m_0c^2$), but rather the expression $H=mv^2$.\\
\\
{\bf 2.2. The experiment of Steinberger et al.\cite{Steinberger}:}\\
\\
The experiment by Steinberger et al. is very similar to the one by Bjorklund et al. It is mentioned here, however, because the experimenters attempted a measurement that was not attempted by the first group: they directly attempted to detect the decay of stationary pions. In that experiment, a target was bombarded with photons (x-rays) instead of protons. A pair of telescopes positioned at variable angles around the target detected the photon pairs resulting from the decay of the neutral pions. In their paper, the authors indicated that they intentionally adjusted the energy of the incident beam to 175Mev, which was just above the production threshold of the pions, and positioned the telescopes at angles of $0^{\circ}$ and $180^{\circ}$. Of course, they reported that the observed photon yield was 50 times smaller than the yield at higher energies. They then concluded that coincidences at $180^{\circ}$ are ``rare". This conclusion, again, is unquestionable proof that the decay of a stationary pion is impossible.\\
\\
\\
{\large\bf 3. Reinterpretation of the charged pion ($\pi^{\pm}$) experiments:}\\
\\
{\bf 3.1. The experiment of Lattes et al.\cite{Lattes} and the questionability of the presence of a neutrino in charged pion decay:}\\
\\
In the 1947 experiment of Lattes et al., the decay of a charged pion into a lighter particle, the muon, was observed for the first time. The authors demonstrated that a highly energetic positive pion from a cosmic shower will be scattered by a nucleus in the photographic emulsion that was exposed to that shower, and, in the process, its mass decreases considerably. The lighter particle was called a ``$\mu$ meson" by the authors. To satisfy Einstein's equation $H=mc^2$, they postulated that a neutrino or another neutral particle must be emitted along with the muon, since the total energy of the muon was obviously considerably less than that of the pion. We shall now proceed to examine this experiment and reach an important conclusion: the data presented by the authors, together with the theory of QCD, do in fact rule out the possibility of the emission of a second particle and further support the total energy equation $H=mv^2$.\\
\\
When we consider a problem such as the one presented in that experiment, the first question that we must ask is whether the scattering of the incoming pion was an elastic or inelastic process. The crucial fact that was presented by the authors was that the resulting muon was always emitted with a constant kinetic energy (about 4Mev), while the incoming pion seemed to have widely different kinetic energies. This fact (whether we accept the idea of the emission of a second particle or not) undoubtedly point to an inelastic collision of the incoming pion, since it indicates that a mass-energy transformation (but no kinetic energy transformation) gives rise to the muon. In other words, this fact indicates that the pion likely loses its kinetic energy almost entirely upon collision with a nucleus in the emulsion, and then a mass-energy transformation process gives rise to another particle with a lower mass and constant kinetic energy: the muon. There are two questions that must now be answered: first, how such a process can be understood within the framework of the QCD theory?; secondly: if the incoming pion loses its kinetic energy entirely (and hence its total energy $H_{\pi} = 0$), how the muon is then generated? We shall answer these two questions in order, and, in the process, we shall reach some very surprising conclusions.\\
\\
According to QCD, the transformation $\pi^{+} \rightarrow \mu^{+}$ can be understood in the following manner\cite{Williams} (note that the positive pion is a quark doublet composed of an up quark and an anti-down quark):

\pagebreak

\vspace*{-0.5in}

\begin{eqnarray}
\lefteqn{\bar{d} \rightarrow \bar{u} + \mu^{+} + \nu} \nonumber\\
\lefteqn{(u+\bar{u}) + \mu^{+} + \nu \rightarrow}  \nonumber\\
 & & \qquad \mbox{Kinetic Energy} +  \mu^{+} + \nu  
\label{311}
\end{eqnarray}

In the first reaction, the anti-down quark $\bar{d}$ decays into an anti-up quark $\bar{u}$, a muon $\mu^{+}$ and a neutrino $\nu$ (we shall shortly demonstrate that the generation of a neutrino is impossible). Note that, generally, this reaction is possible if the mass of the $\bar{d}$ quark is taken to be the relativistic mass, not the rest mass (which is known to be about 5Mev to 8Mev \cite{PDG}). In the second reaction, the up quark annihilates with its anti-particle ($\bar{u}$) and the resulting energy appears as kinetic energy of the particles that were generated in the first reaction. 
In the earlier paper by the author\cite{Bakhoum1} it was demonstrated that in the case of the annihilation of a particle with its anti-particle it is reasonable to assume that a quantity of energy equal to $2m_0 c^2$ will be obtained. We now note that the quantity $2m_0 c^2$, where $m_0$ is the rest mass of the up quark, is approximately 3 to 9Mev according to the latest theoretical estimates\cite{PDG}. This figure, of course, agrees very well with the muon kinetic energy of 4Mev given by Lattes et al. ! This result suggests that it is very doubtful that a neutrino is generated along with the muon in the process of pion decay, especially since the authors estimated that the neutrino must carry away an energy of approximately 29Mev in order to balance the relativistic mass-energy of the pion! Quite clearly, the expected Q-value of the reaction is in agreement with the measured kinetic energy of the muon, without the need for the neutrino hypothesis.\\
\\
We next consider the question of how the muon is generated and why this data does in fact support the total energy equation $H=mv^2$. If we calculate the velocity $v$ of the resulting muon, using its known rest mass (105.66Mev/$c^2$), and based on a kinetic energy of 4Mev, we find that $v \approx 0.27 c$ (i.e., sub-relativistic). At sub-relativistic velocities, the total energy $H=mv^2$ is twice the kinetic energy; hence $H_{\mu} \approx 8$ Mev. But if we reach the conclusion that $H_{\pi} = 0$, then how the muon is generated? This process obviously occurs in the vicinity of a nucleus in the emulsion, as a result of the interaction of the electrostatic field of the nucleus with the constituent quarks in the pion. If we look at the overall mass-energy transformation equation, this equation can be written as follows

\begin{equation}
u \bar{d} \rightarrow H_{\mu} \: (=8 \mbox{Mev})
\label{312}
\end{equation}

The two quarks, of course, being charged, will acquire kinetic and potential energy terms as a result of the presence of a nearby electrostatic field. As we indicated, since we expect the annihilation of a particle and an antiparticle in this process, then it is reasonable to assume that an amount of energy equal to $m_0c^2$ will be obtained for each of the particles involved in the reaction (see the discussion in ref.\cite{Bakhoum1}). The expected mass-energy of the two quarks in Eq.(\ref{312}) will be therefore given by the sum $m_0c^2 (u) + m_0 c^2 (d)$. This quantity is approximately 6.5 to 13Mev, according to the latest estimates\cite{PDG}. It is now clear that this result supports the prediction of a total muon energy ($H_{\mu}$)  of 8Mev, in agreement with the total energy equation $H=mv^2$! Of course, it is also clear from this discussion that one must proceed with great caution in interpreting particle reactions, since the traditional quantity of energy $mc^2$ may indeed appear in certain cases as we have seen.\\
\\
To summarize the above discussion: the theory of QCD, together with the data presented by Lattes et al., rule out the possibility of the emission of a neutrino and further support the total energy equation $H=mv^2$. On the other hand, the application of the total energy equation $H=mc^2$ results in an energy deficit of 29Mev (or 34Mev according to the modern estimates) that cannot be accounted for.\\
\\
{\bf 3.2. The experiment of Danby et al.\cite{Lederman2} and the questionability of the existence of a ``muon neutrino":}\\
\\
The 1962 experiment of Danby et al. (which Lederman described in his book ``The God Particle") was designed to detect the presence of neutrinos in the decay of pions and kaons. In that experiment, pions, kaons and neutrons (products of the collision of 15Gev protons with a beryllium target) traveled through an unshielded distance of 21 meters, followed by another distance of 13.5 meters through a steel shield. Any particle that survived the path through the steel shield finally entered a carefully designed spark chamber. The essential requirement in the experiment (designed to detect the neutrinos only) was that a meaningful ``event" must be one that originates inside the chamber. In other words, to reject muons or other charged particles that survive the steel shield, it was necessary that an event that counts must be one that is initiated by a neutral particle that penetrates the chamber. Such a neutral particle would then impact the aluminum plates inside the chamber, generating an energetic electron or muon. This objective was achieved by counting only those events in which the first two spark gaps in the chamber did not ignite, indicating that the event was induced by a neutral particle. Of course, the primary candidate for the cause of those observed events was the neutrino. However, since most of those events resulted in an energetic muon, Lederman and his colleagues reached the conclusion that the neutrino in this case must be a neutrino of a different \linebreak ``flavor": a ``muon neutrino".\\
\\
In view of the above discussion of the Lattes et al. experiment, how can we interpret Lederman's experiment if we reach the conclusion that pion decay does not result in a neutrino? There are two potential explanations for the events that were observed in that experiment (regrettably, these were not considered by the experimenters):\\
\\
1. A decay sequence such as\cite{Appel}

\begin{eqnarray}
K^{+} & \rightarrow & \pi^{+} \pi^{0} \nonumber\\
\pi^{0} & \rightarrow & \mu^{-} e^{+}
\label{}
\end{eqnarray}

In that decay sequence, a charged kaon decays into a charged pion and a neutral pion. Since the mean lifetime of the charged kaon is on the order of $10^{-8}$ sec., and since relativistic time dilation should be expected (given the relativistic velocities of the particles as they emerge from the target), we can see that the charged kaon can very reasonably be expected to travel a major distance through the steel shield before it decays. In the second reaction one of the decay products, the neutral pion, penetrates the chamber, but decays only after it passes the first two spark gaps. The resulting positron is then immediately absorbed in the aluminum plates, while the muon continues through the gaps and hence gets detected as an ``event". Now we note that the first decay reaction has a branching ratio of about 21\%. The second reaction, however, has a branching ratio on the order of $10^{-8}$ (for the combined $\mu^{-} e^{+} + \mu^{+} e^{-}$ decays); but this is actually much higher than the percentage of events observed in the experiment and attributed to the neutrino. Understandably, the 1962 experimenters believed that reactions that violate the empirical law of lepton number conservation (such as this one) cannot occur; but recent evidence suggests otherwise\cite{Appel}.\\
\\
2. Danby et al. convincingly ruled out the possibility that the cause of the events is the neutrons that originated from the target; they also convincingly demonstrated that the cause must be the pions and the kaons. However, the possibility of ``delayed" neutrons was not considered. Delayed neutrons can be generated along the flight path of the pions, by means of the reaction\cite{Williams}

\begin{equation}
\pi^{+} + p \rightarrow \pi^{+} + \pi^{+} + n
\label{}
\end{equation}

In this reaction, a charged pion travels a certain distance through the steel shield, strikes a proton and gives rise to a delayed neutron. Those delayed neutrons can then cause the events observed in the spark chamber (note that the decay of the charged pions is what gave rise to the large numbers of charged muons that successfully penetrated the shield, as documented by the experimenters. This leaves little doubt that delayed neutrons that share the same origin will also penetrate the shield).\\
\\
In summary, if we rule out the possibility of the existence of a ``muon neutrino", then there are other potentially plausible explanations for the events observed by Lederman and his colleagues in the famous 1962 experiment.\\ 
\\
\\
{\large\bf 4. Summary and additional conclusions:}\\
\\
Since the introduction of the neutrino hypothesis by Pauli in 1930, the neutrino became the standard vehicle that is relied upon to compensate for the failure of the equation $H=mc^2$ to explain the majority of the experimental facts. In view of this paper and the earlier paper by the author\cite{Bakhoum1}, the neutrino must be understood within a very limited context: it is likely a highly penetrating particle that emerges in certain high-energy nuclear reactions, such as solar reactions, but it is not Pauli's particle and it should not be used as the standard ``solution" when the equation $H=mc^2$ fails to explain the experimental results. As we have seen, such failure has led to a path of enormous difficulties and has necessitated the introduction of exotic concepts like ``neutrino flavors" and ``flavor oscillations". It is indeed regrettable that particle physics has reached such a sad state of affairs.\\
\\
To terminate our discussion, we show in Fig.2 below the spectrum of the electron's kinetic energy that is typically obtained in muon decay experiments\cite{Perkins},\cite{Goldhaber}:

\par
\vspace*{.15in}

\psfig{figure=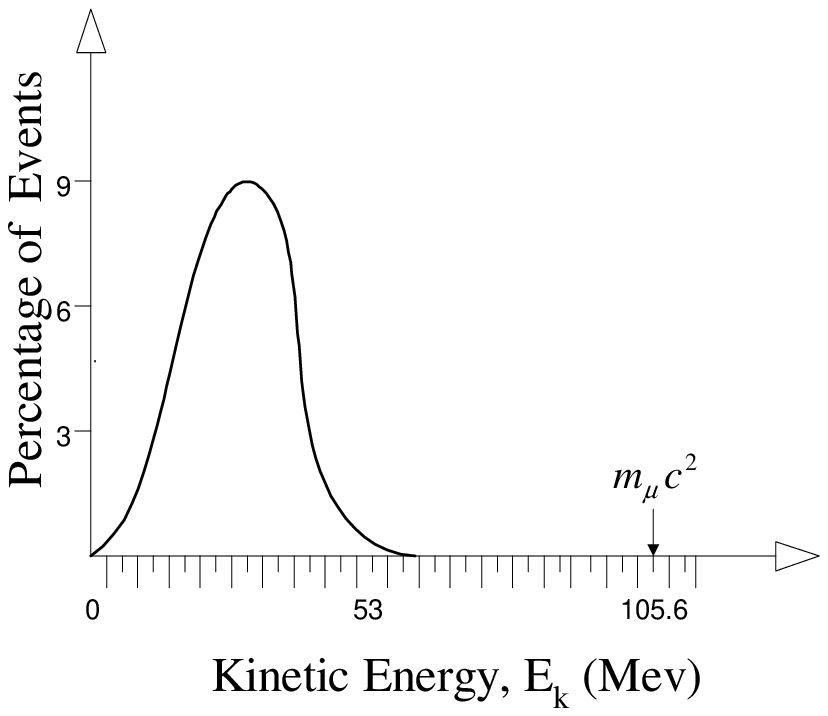}

\vspace*{.05in}

{\small Figure 2: Typical plot of the distribution of the electron's (or positron's) kinetic energy in muon decay experiments. Like many other cases in particle physics, the expected quantity of energy $m_{\mu} c^2$ (105.6Mev) fails to emerge. The distribution can be understood in terms of the quantity $m_{\mu} v^2_{\mu}$, rather than the hypothesis of $m_{\mu} c^2$ minus two neutrinos.}

\pagebreak

As we can observe from Fig.2, the spectrum of the kinetic energy of the electron or positron that emerges from the decay of a muon is a continuous spectrum, just like many other cases in particle physics. In view of the total energy equation $H=mv^2$, this distribution can be understood as follows: the principle of conservation of energy implies that $mv^2$(muon) = $mv^2$(electron). But since the electron emerges with a relativistic velocity, then $mv^2$(electron) is approximately equal to its kinetic energy $E_k$ (since the mass-energy of 0.511Mev is negligible). Moreover, given the fact that the muon usually decays while traveling at sub-relativistic velocities, then we have

\begin{equation}
E_k \approx m_{\mu} v^2_{\mu},
\label{}
\end{equation}

where $m_{\mu}$ is the rest mass of the muon. Since the velocity of the decaying muon can vary widely, we can now clearly see that this expression explains perfectly the distribution in Fig.2. In addition, it is now obvious why a reaction such as 

\begin{equation}
\mu^{+} \rightarrow e^{+} + \gamma
\label{}
\end{equation}

has never been observed experimentally. There is simply no ``lost energy" in muon decay that must appear in the form of a photon or a neutrino\footnote{Note that since muons decay in matter, then the principle of momentum conservation implies that the momentum of a nearby nucleus must be factored in the momentum balance. The measured kinetic energy of the electron, however, will be nearly equal to the total energy of the reaction due to the electron's tiny mass.}. Wouldn't the total energy equation $H=mv^2$ be a simpler explanation than the hypothesis of ``a muon decays into an electron plus a neutrino plus an anti-neutrino"? One only has to look at a figure like Fig.2 and wonder why it took more than 70 years to reach the conclusion that $H=mc^2$ is an inaccurate physical law.\\
\\
\\
{\large\bf Acknowledgements:}\\
\\
The author is grateful to Dr. Walter Becker, retired, for valuable suggestions on improving the paper. The author also thanks Dr. Christopher Peters of the New Jersey Institute of Technology for many useful discussions about the issues presented.


\begin{thebibliography}{99}

\bibitem{Bakhoum1} E. Bakhoum, {\sl Fundamental Disagreement of Wave Mechanics with Relativity\/}, Physics Essays, 15, 1, 2002. Online e-print archive: physics/0206061.

\bibitem{Lederman1} L. Lederman, {\sl The God Particle\/}, (Hought\-on Mifflin, New York, NY, 1993), p.293.

\bibitem{Lederman2} G. Danby et al., {\sl Observation of High Energy Neutrino Reactions and the Existence of Two Kinds of Neutrinos\/}, Phys. Rev. Lett., 9, 1, p.36 (1962). Reprinted in ref.\cite{Goldhaber}.

\bibitem{Lattes} C.M.G. Lattes et al., {\sl Observations on the Tracks of Slow Mesons in Photographic Emulsions\/}, Nature, 160, p.453 (1947). Reprinted in ref.\cite{Goldhaber}.

\bibitem{Bjorklund} R. Bjorklund et al., {\sl High Energy Photons from Proton-Nucleus Collisions\/}, Phys. Rev., 77, 2, p.213 (1950).

\bibitem{Steinberger} J. Steinberger et al., {\sl Evidence for the Production of Neutral Mesons by Photons\/}, Phys. Rev., 78, p.802 (1950). Reprinted in ref.\cite{Goldhaber}.

\bibitem{Appel} R. Appel et al., {\sl An Improved Limit on the Rate of the Decay $K^{+} \rightarrow \pi^{+} \mu^{+} e^{-}$\/}, Phys. Rev. Lett., 85, p.2450 (2000).

\bibitem{Williams} W.S.C. Williams, {\sl Nuclear and Particle Physics\/} (Oxford Univ. Press, 1991).

\bibitem{PDG} Particle Data Group, http://pdg.lbl.gov, 2002.

\bibitem{Perkins} D.H. Perkins, {\sl Introduction to High Energy Physics\/} (Addison-Wesley, Reading, MA, 1987).

\bibitem{Goldhaber} R.N. Cahn and G. Goldhaber, {\sl The Experimental Foundations of Particle Physics\/} (Cambridge Univ. Press, 1989).




\end{thebibliography}
\end{document}